\documentclass[a4paper]{jpconf}
\usepackage{graphicx}

\begin{document}
\title{Universal Property of Quantum Gravity implied by Bekenstein-Hawking Entropy and Boltzmann formula}

\author{Hiromi Saida}

\address{Department of Physics, Daido University, Nagoya 457-8530, Japan}

\ead{saida@daido-it.ac.jp}

\begin{abstract}
We search for a universal property of quantum gravity. 
By ``universal'', we mean the independence from any existing model of quantum gravity (such as the super string theory, loop quantum gravity, causal dynamical triangulation, and so on). 
To do so, we try to put the basis of our discussion on theories established by some experiments. 
Thus, we focus our attention on thermodynamical and statistical-mechanical basis of the black hole thermodynamics: 
Let us assume that the Bekenstein-Hawking entropy is given by the Boltzmann formula applied to the underlying theory of quantum gravity. 
Under this assumption, the conditions justifying Boltzmann formula together with uniqueness of Bekenstein-Hawking entropy imply a reasonable universal property of quantum gravity. 
The universal property indicates a repulsive gravity at Planck length scale, otherwise stationary black holes can not be regarded as thermal equilibrium states of gravity. 
Further, in semi-classical level, we discuss a possible correction of Einstein equation which generates repulsive gravity at Planck length scale. 
\end{abstract}

%%%%%%%%%%%%%%%%%%%%%%%%%%%%%%%%%%%%%%%%%%%%%%%%%%%%%%%%%%%%%%%%%%%%%%%%%%%%%%%%%%%%%%%%%%%%%%%%%%%%
\section{Introduction}
\label{O28-13:sec:intro}

We investigate the basis of black hole thermodynamics in order to search for universal properties of quantum gravity, where ``universal'' means the independence from any existing model of quantum gravity. 
The basic concept of black hole thermodynamics is as follows: 
\emph{Because a stationary black hole emits the thermal radiation (Hawking radiation), the black hole is in a thermal equilibrium state of gravity.
And, usually, the micro-scopic origin of black hole is expected to be the quantum states of gravity.}
This concept depends on the thermal radiation theory which indicates that the source of ``Planck spectrum'' is a thermal equilibrium object.

By accepting the black hole thermodynamics, it is reasonable to require that the quantum gravity satisfies the presupposition of thermal radiation theory, otherwise we can not regard black holes as thermal equilibrium states by observing the Hawking radiation. 
Here, the presupposition of thermal radiation theory is that the ordinary thermodynamics and quantum statistical mechanics work well on any thermal equilibrium object (at least in laboratory system). 
Therefore, it is natural to require that \emph{the basic idea of statistical mechanics is applicable to thermal equilibrium states of not only laboratory systems but also quantum gravity (black holes)}. 
Here, the basic idea of statistical mechanics is, for example, the Boltzmann formula, $S_{BH} = \ln\Omega_{BH}$, where $S_{BH}$ is the Bekenstein-Hawking entropy and $\Omega_{BH}$ is the number of states of black hole counted by the underlying quantum gravity.

From the above, we adopt the following two suppositions for our purpose: 
\begin{description}
\item[\bf Supposition 1:]
 A stationary black hole is a thermal equilibrium state composed of micro-states of underlying quantum gravity. 
\item[\bf Supposition 2:]
 Basic formalism of statistical mechanics (e.g. Boltzmann formula, grand partition function and so on) is applicable to underlying quantum gravity so as to describe the thermal equilibrium states (black holes). 
\end{description}
The issues for each supposition are as follows:
\begin{list}{}{}
\item[\bf Issue on Supposition 1 (Uniqueness of BH entropy):] 
Note that, if the Boltzmann formula, $S_{BH} = \ln\Omega_{BH}$, holds for black holes, then it defines the Bekenstein-Hawking entropy $S_{BH}$ \emph{uniquely} in micro-scopic point of view. 
This means that the Boltzmann formula can never work well unless the uniqueness of entropy is proven in macro-scopic point of view of thermodynamics (independent of any micro-scopic theory)~\cite{O28-13:ref:lieb+1.1999,O28-13:ref:tasaki.2000,O28-13:ref:tasaki.2008}. 
In ordinary thermodynamics, the uniqueness of entropy is proven by some basic axioms of thermodynamics~\cite{O28-13:ref:lieb+1.1999,O28-13:ref:tasaki.2000}. 
However, a basic axiom of ordinary thermodynamics do not hold in black hole thermodynamics~\cite{O28-13:ref:saida.2011,O28-13:ref:york.1986,O28-13:ref:saida.2010}. 
Then, it is not necessarily manifest whether the entropy $S_{BH}$ is unique or not in black hole thermodynamics. 
Hence, we have to prove the uniqueness of $S_{BH}$ before the discussion of Boltzmann formula (see Sec.\ref{O28-13:sec:entropy}).
\item[\bf Issue on Supposition 2 (Justification of Boltzmann formula):] 
Once the uniqueness of $S_{BH}$ is proven in macro-scopic point of view, we can proceed to discussions on Boltzmann formula. 
Note that the validity of Boltzmann formula in ordinary quantum statistical mechanics is justified by some intrinsic properties of Hilbert space of ordinary quantum mechanics. 
Those intrinsic properties appear in Ruelle-Tasaki theorem and Dobrushin theorem~\cite{O28-13:ref:tasaki.2008,O28-13:ref:ruelle.1999,O28-13:ref:dobrushin.1964} (see Sec.\ref{O28-13:sec:boltzmann}). 
\emph{Thus, if $S_{BH}$ is given by the Boltzmann formula, we can expect that those theorems of ordinary quantum mechanics are related with some property of underlying quantum gravity. 
Some universal property of quantum gravity may be extracted from those theorems} (see Sec.\ref{O28-13:sec:conc}).
\end{list}

%%%%%%%%%%%%%%%%%%%%%%%%%%%%%%%%%%%%%%%%%%%%%%%%%%%%%%%%%%%%%%%%%%%%%%%%%%%%%%%%%%%%%%%%%%%%%%%%%%%%
\section{Uniqueness Theorem of Bekenstein-Hawking Entropy}
\label{O28-13:sec:entropy}

In the rigorous axiomatic formulation of ordinary thermodynamics~\cite{O28-13:ref:lieb+1.1999,O28-13:ref:tasaki.2000}, the basic axioms consist of not only the ``four laws of thermodynamics'' but also some more requirements. 
One of the axioms of ordinary thermodynamics modified in black hole thermodynamics is the classification requirement of state variables; all thermodynamic state variables are classified into two categories, extensive variables (e.g. energy, entropy and so on) and intensive variables (e.g. temperature, pressure and so on). 
However, as explained in detail in ref.\cite{O28-13:ref:saida.2011,O28-13:ref:saida.2010}, the state variables in black hole thermodynamics can not be classified into the two categories, but classified into \emph{three} categories. 
Hence, various theorems (including the uniqueness of entropy) in ordinary thermodynamics, whose proof uses the classification of state variables, should be re-proven in black hole thermodynamics. 
Ref.\cite{O28-13:ref:saida.2011} gives an axiomatic proof to the uniqueness of $S_{BH}$:
\begin{description}
\item[Uniqueness Theorem of Entropy:] 
Let $K$ be an extensive variable which increases along irreversible adiabatic processes. 
Then there exist two constants $a\,(>0)$ and $b$ such that $K = a S_{BH} + b$, where $S_{BH}$ is the Bekenstein-Hawking entropy. 
(See ref.\cite{O28-13:ref:saida.2011} for the proof.)
\end{description}
Once this theorem is proven, we can proceed to discussions on Boltzmann formula.
In this sense, we can regard the conclusion in Sec.\ref{O28-13:sec:conc} as a result implied by the uniqueness of $S_{BH}$.

%%%%%%%%%%%%%%%%%%%%%%%%%%%%%%%%%%%%%%%%%%%%%%%%%%%%%%%%%%%%%%%%%%%%%%%%%%%%%%%%%%%%%%%%%%%%%%%%%%%%
\section{Conditions Justifying Boltzmann Formula}
\label{O28-13:sec:boltzmann}

Consider the ordinary quantum mechanics in this section. 
For simplicity, consider the system of $N$ identical particles in a region of volume $V$ with no external field, and:
\begin{list}{}{}
\item[$\bullet$ Interaction potential :]
$\Phi(\vec{x}_1,\cdots,\vec{x}_N) = \sum_{j=1}^{\infty}\phi^{(j)}(\vec{x}_{i_1},\cdots,\vec{x}_{i_j})$\,, where $i_j \in (1, \cdots , N)$\,.
\item[$\bullet$ Energy eigen value :]
$E_k(V,N)$\,, where $k = 0, 1, 2, \cdots$ and $E_k \le E_{k+1}$ 
(``$=$'' for degeneracy).
\item[$\bullet$ Number of states :]
$\Omega(V,N;U) :=$ ``Number of eigen states satisfying $E_k \le U$'' $= \max\limits_{E_k\le U} k$ \,.
\end{list}
For the first, let us show the Ruelle-Tasaki theorem which clarifies the sufficient conditions for the validity of Boltzmann formula~\cite{O28-13:ref:saida.2011,O28-13:ref:ruelle.1999,O28-13:ref:tasaki.2008}:
\begin{description}
\item[Ruelle-Tasaki Theorem:]
Suppose the following two conditions of $\Phi(\vec{x}_1,\cdots,\vec{x}_N)$:
\begin{description}
\item[Condition~A:]
Arbitrary $j$-particle interaction, $\phi^{(j)}$, becomes negative for sufficiently large distribution of $j$ particles.
That is, there exists a constant $r_A\,(>0)$, such that\, 
$\phi^{(j)}(\vec{x}_{i_1},\cdots,\vec{x}_{i_j}) \le 0$ 
\,\, for \,\,
$r_A \le \min\limits_{k, l = 1,\cdots,j}\left| \vec{x}_{i_k} - \vec{x}_{i_l} \right|$
\,.
\item[Condition~B:]
Interaction potential $\Phi$ is bounded below.
That is, there exists a constant $\phi_B\,(>0)$, such that\, 
$\Phi(\vec{x}_1,\cdots,\vec{x}_N) \ge -N\,\phi_B$ \,.
\end{description}
Then, the limit, 
$\displaystyle 
\sigma(\varepsilon,\rho) := 
 \lim_{t.l.}\frac{\ln\Omega(V,N;U)}{V}$, 
exists uniquely, where $\lim\limits_{t.l.}$ is thermodynamic limit defined by $V\to\infty$ with fixing $\rho := N/V$ and $\varepsilon := U/V$ at constant values.
\end{description}
By this theorem, the thermodynamics limit of $\ln\Omega(V,N;U)$ is defined well. 
Furthermore, it can be also proven that $\sigma(\varepsilon,\rho)$ is concave about its arguments $(\varepsilon,\rho)$ and monotone increasing about $\varepsilon$, and remains constant along \emph{reversible} adiabatic processes~\cite{O28-13:ref:tasaki.2008}. 
These behaviors are some characteristic properties of entropy already known in thermodynamics. 
Then, in statistical mechanics, it is assumed that $\ln\Omega(V,N;U)$ is equal to the entropy (the Boltzmann formula).

Obviously, the above conditions~A and~B are the sufficient conditions for the validity of Boltzmann formula. 
Furthermore, a system, which holds Boltzmann formula but violates conditions~A or~B, has not been found experimentally. 
\emph{Hence, at least in laboratory systems, it is reasonable to require that any physical system satisfies the conditions~A and~B.}

Next, let us show the Dobrushin theorem which clarifies the necessary condition for the existence of thermal equilibrium states~\cite{O28-13:ref:saida.2011,O28-13:ref:ruelle.1999,O28-13:ref:dobrushin.1964}:
\begin{description}
\item[Dobrushin theorem:] 
Consider the case satisfying following two presuppositions:
\begin{description}
\item[Presupposition~C:]
The $j(\neq 2)$-particle interactions disappear, and the total interaction potential is a sum of two-particle interactions, $\Phi(\vec{x}_1,\cdots,\vec{x}_N) = \sum\limits_{1\le i < j \le N} \phi^{(2)}(\vec{x}_i,\vec{x}_j)$.
\item[Presupposition~D:]
$\phi^{(2)}(\vec{x},\vec{y}) \propto |\vec{x}-\vec{y}|^{-\alpha}$ ($\alpha > 0$) for sufficiently large $|\vec{x}-\vec{y}|$.
\end{description}
Under these presuppositions, if the ground partition function can be defined uniquely (i.e. if thermal equilibrium states exist), then the following inequality holds,
\begin{equation}
\label{eq:bf.thm.d.IV}
 I_V^{(2)} :=
 \frac{1}{V^2}\int\int_V {\rm d}^3x_1\,{\rm d}^3x_2\, \phi^{(2)}(\vec{x}_1,\vec{x}_2) \ge 0 \,.
\end{equation}
\end{description}
Obviously, $I_V^{(2)} \ge 0$ is the necessary condition for the existence of thermal equilibrium states. 
Note that the presuppositions~C and~D are natural when we consider the gravity.

%%%%%%%%%%%%%%%%%%%%%%%%%%%%%%%%%%%%%%%%%%%%%%%%%%%%%%%%%%%%%%%%%%%%%%%%%%%%%%%%%%%%%%%%%%%%%%%%%%%%
\section{Conclusion and Discussion}
\label{O28-13:sec:conc}

\begin{figure}[h]
\includegraphics[width=65mm]{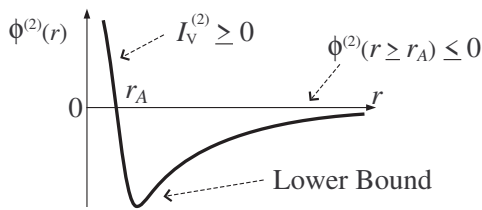}\hspace{5mm}%
\begin{minipage}[b]{85mm}\caption{\label{fig:1}Typical form of the effective potential of quantum gravity at semi-classical level. Thermal equilibrium states of gravity (black holes) may not exist unless a repulsive region appears, where $r_A$ may be Planck scale.}
\end{minipage}
\end{figure}

Our supposition~2 in Sec.\ref{O28-13:sec:intro} implies that the ordinary quantum mechanics and quantum gravity share the same properties which justify the Boltzmann formula and existence of thermal equilibrium states. 
The sufficient conditions~A, B and the necessary condition $I_V^{(2)} \ge 0$ are expected to be such shared properties~\footnote{ 
So far, as a fact, no counter-example to conditions~A and~B are found experimentally. 
It may be reasonable to extend this fact to quantum gravitating systems, so that quantum gravity also satisfies those conditions.
}. 
The typical form of the potential is shown in Fig.\ref{fig:1}. 
Note that, these three conditions (A, B and $I_V^{(2)}\ge 0$) are for the interaction potential.

On the other hand, it is not clear whether the full quantum gravity is described by using the interaction potential $\Phi$ or Hamiltonian (or Lagrangian). 
However, it seems to be certain that a semi-classical expression of quantum gravity is described by an effective field equation of spacetime metric (with some auxiliary field if needed). 
Hence, our suggestion is as follows: 

\begin{description}
\item[Implication from the above:]
If the effective field equation of quantum gravity at semi-classical level is described by a correction term $f_{\mu\nu}$ added to Einstein equation such as
\begin{equation}
\label{eq:conc.eom}
\displaystyle 
 R_{\mu\nu} -\frac{1}{2}R\,g_{\mu\nu} + f_{\mu\nu} = \frac{8\pi G}{c^4} T_{\mu\nu} \,,
\end{equation}
(where $T_{\mu\nu}$ is the energy-momentum tensor, $g_{\mu\nu}$ the metric, $R_{\mu\nu}$ the Ricci tensor, and $R := R_{\mu}\,^{\!\!\mu}$), 
then $f_{\mu\nu}$ should generate the repulsive gravity at short (e.g. Planck) length scale, and a necessary property of $f_{\mu\nu}$ is given in Eq.(\ref{eq:conc.f}) below.
\end{description}
Let us note that the Raychaudhuri equation holds for arbitrary timelike/null congruence in arbitrary spacetime. 
For simplicity, consider the local hypersurface orthogonal (i.e. vanishing vorticity) geodesic congruence. 
Then, the Raychaudhuri equation becomes, 
${\rm d}\theta/{\rm d}\lambda =
 -R_{\mu\nu}u^{\nu}u^{\nu} -\sigma^2 - \theta^2/p$, 
where $\lambda$ is the affine parameter, $u^{\mu}$ the tangent of geodesic, $\theta := u^{\mu}\,_{\!;\mu}$ is the expansion (the rate of change of sectional area of congruence, d[area]$/{\rm d}\lambda$), $\sigma^2$ the square of shear tensor, and $p = 3$ or $2$ for respectively timelike or null geodesics. 
Note that, in the general relativity ($f_{\mu\nu} = 0$), the so-called \emph{energy condition} (e.g. the null energy condition, $T_{\mu\nu}u^{\mu}u^{\nu} \ge 0$ for arbitrary null vector $u^{\mu}$) gives $-R_{\mu\nu}u^{\mu}u^{\nu} \le 0$ by the Einstein equation and results in ${\rm d}\theta/{\rm d}\lambda \le 0$ by the Raychaudhuri equation, which indicates the attractive nature of gravity. 
On one hand, if there arises a case of $-R_{\mu\nu}u^{\nu}u^{\nu} > 0$, then ${\rm d}\theta/{\rm d}\lambda > 0$ may hold by the Raychaudhuri equation, which is regarded as a necessary condition of repulsive gravity.

Then, when we require the repulsive gravity in Planck length scale, the correction term $f_{\mu\nu}$ should give the inequality $-R_{\mu\nu}u^{\nu}u^{\nu} > 0$ in Planck length scale. 
Here, by Eq.(\ref{eq:conc.eom}), we find, 
$-R_{\mu\nu}u^{\nu}u^{\nu} =
 -(8\pi G/c^4)\,\bigl[\,T_{\mu\nu}u^{\mu}u^{\nu} - (1/2) T\,u^2 \,\bigr]
 +\bigl[\, f_{\mu\nu}u^{\mu}u^{\nu} - (1/2) f\,u^2 \,\bigr]$, 
where $T = T_{\mu}\,^{\!\mu}$ and $f = f_{\mu}\,^{\!\mu}$. 
Hence, the inequality $-R_{\mu\nu}u^{\nu}u^{\nu} > 0$ turns to be,
\begin{equation}
\label{eq:conc.f}
\mbox{\bf Necessary property in Planck scale:}\quad
 f_{\mu\nu}u^{\mu}u^{\nu} - \frac{1}{2} f\,u^2 \,\,>\,\,
 \frac{8\pi G}{c^4}\Bigl[T_{\mu\nu}u^{\mu}u^{\nu} - \frac{1}{2} T\,u^2 \Bigr]
\end{equation}
Under Eq.(\ref{eq:conc.f}), a suitable form of $f_{\mu\nu}$ may be possible, which raises a repulsive gravity with preserving traditional energy conditions (e.g. dominant, weak and strong energy conditions).

In this manuscript, any existing model of quantum gravity is not used. 
(See ref.\cite{O28-13:ref:saida.2011} for detail.) 
Therefore, the above implication can be regarded as a universal property of quantum gravity.

%%%   References

\end{document}